\documentclass{aa}
\usepackage{epsfig}
\def\ltsima{$\; \buildrel < \over \sim \;$}
\def\simlt{\lower.5ex\hbox{\ltsima}}
\def\gtsima{$\; \buildrel > \over \sim \;$}
\def\simgt{\lower.5ex\hbox{\gtsima}}
\begin{document}
\thesaurus{}

   \thesaurus{6(02.01.2; 
	   02.12.1;  
	   02.12.3;  
	   11.09.1: MCG-6-30-15;  
	   11.19.1;  
	   13.25.3)} 

\title{BeppoSAX confirms extreme relativistic effects in the X-ray spectrum of MCG-6-30-15}

\author{M. Guainazzi\inst{1}, G. Matt\inst{2}, S. Molendi\inst{3}, A. Orr\inst{1}, F. Fiore\inst{4,5}, P. Grandi\inst{6}, A. Matteuzzi\inst{5}, T. Mineo\inst{7}, G.C. Perola\inst{2}, A.N. Parmar\inst{1}, L. Piro\inst{6}}

\institute{
{Astrophysics Division, Space Science Department of ESA, ESTEC, Postbus 299,
NL-2200 AG Noordwijk, The Netherlands}
\and
{Dipartimento di Fisica ``E.Amaldi'', Universit\`a degli Studi ``Roma Tre'', Via della Vasca Navale 84, I-00146 Roma, Italy}
\and
{Istituto di Fisica Cosmica ``G.Occhialini'', CNR, Via Bassini 15, I-20133 Milano, Italy}
\and
{Osservatorio Astronomico di Roma, Via dell'Osservatorio, I-00144 Monteporzio Catone, Italy}
\and
{BeppoSAX Science Data Center, Via Corcolle 19, I-00131 Roma, Italy}
\and
{Istituto di Astrofisica Spaziale CNR, Via Fosso del Cavaliere, I-00133 Roma, Italy}
\and
{Istituto di Fisica Cosmica ed Applicazioni dell'Informatica CNR, Via Ugo La Malfa 153, I-90146 Palermo, Italy}
}
   
\offprints{M.Guainazzi [mguainaz@astro.estec.esa.nl]}

\date{Received;accepted}

\maketitle

\markboth{M.Guainazzi et al.}{BeppoSAX confirms extreme relativistic effects in the X-ray spectrum of MCG-6-30-15}

\begin{abstract}

We report in this {\it Letter} the first simultaneous measure of
the X-ray broadband (0.1--200~keV) continuum and of the iron K$_{\alpha}$
fluorescent line profile in the Seyfert~1 galaxy MCG-6-30-15.
Our data confirms the ASCA detection of a skewed and
redshifted line profile (Tanaka et al. 1995).
The most straightforward explanation is that the line photons are
emitted in the innermost regions of a X-ray illuminated relativistic disk.
The line Equivalent Width (${\rm \simeq 200}$~eV) is perfectly consistent
with the expected value for solar abundances, given the observed amount of
Compton reflection. We report also the discovery of a cut-off in
the nuclear primary emission at the energy of $\simeq$160~keV.

\end{abstract}


\section{Introduction}

There is general consensus on the idea that the energy
output of Active Galactic Nuclei (AGN) is due to the release of radiation
by matter falling onto a supermassive black hole. Evidence of the
presence of such a black hole was only indirect before the launch
of the X-ray satellite ASCA. A long-look observation of
the Seyfert~1 galaxy MCG-6-30-15 revealed that the profile
of the K$_{\alpha}$ iron fluorescent line is broad, skewed and
double-horned (Tanaka et al. 1995).
These are the expected signatures of kinematics and relativistic
effects when photons are emitted
in the innermost regions of a X-ray illuminated accretion disk (Fabian et al. 1989; Matt et al.
1992). Alternative explanations for the observed line
turn out to be unplausible (Fabian et al. 1995).
The iron line is generally broad in the
Seyfert 1s observed by ASCA
(Mushotzky et al. 1995; Guainazzi et al. 1997; Nandra et al. 1997). However,
the limited ASCA energy bandpass did not allow a precise determination of
the underlying continuum, and - in particular - of the
reflection component. Usually, reflection from an infinite, plane-parallel
slab was assumed, following the findings of the X-ray satellite Ginga
(Nandra \& Pounds 1994). This can in principle affect significantly
the measurement of the iron line profile, as shown by Cappi et al.
(1996).

The scientific payload onboard BeppoSAX (Boella et al. 1997a) covers
for the first time simultaneously the broad energy band between 0.1 and
200~keV. It is therefore particularly suited to properly
deconvolve complex spectra. Although the energy resolution of the BeppoSAX
detectors is worse than that of ASCA CCD's ($\simeq 8\%$ versus $\simeq 2\%$ at
6~keV), the better estimate of the underlying continuum can provide 
complementary information on the broad iron lines.
In this {\it Letter} we report on the BeppoSAX study of
the iron line in MCG-6-30-15, which is both the best case for a relativistic
line in ASCA data and one of the Seyfert 1s observed by BeppoSAX with a
long exposure time ($\sim 160$~ks). In the following we will focus
on the high energy, time-average spectrum. The soft X-ray spectrum
is described in detail by Orr et al. (1997).

\section{Observation and data preparation}

MCG-6-30-15 was observed by BeppoSAX from 1996 July 29, 18:49:48 UT to August
3 03:15:00. Data reduction and  analysis follow
the guidelines in Matt et al. (1997).
Only data from the Low Energy Concentrator Spectrometer
(LECS, 0.1--4~keV; Parmar et al. 1997), the Medium Energy Concentrator
Spectrometer (MECS; 1.8--10.5~keV; Boella et al. 1997b) and the
Phoswitch Detector System (PDS; 13--200~keV; Frontera et al. 1997) are
discussed. 
Total exposure times were about
42~ks, 164~ks,
and 86~ks for the LECS, MECS and PDS,
respectively.
Spectra of the imaging instruments have been extracted from circular regions
of radius 8' around the centroid of the source. Total net count rates
were $0.474 \pm 0.003$~s$^{-1}$, $0.966 \pm 0.002$~s$^{-1}$ and
$0.70 \pm 0.03$~s$^{-1}$ in the LECS, MECS and PDS, respectively.

In this {\it Letter}: errors are quoted at 90\% level of confidence for each
parameter (${\rm \Delta \chi^2 = 2.71}$,
Lampton, Margon \& Bowyer 1976); energies are in the source rest frame.

\section{Results}

\begin{figure}
\begin{center}
\epsfig{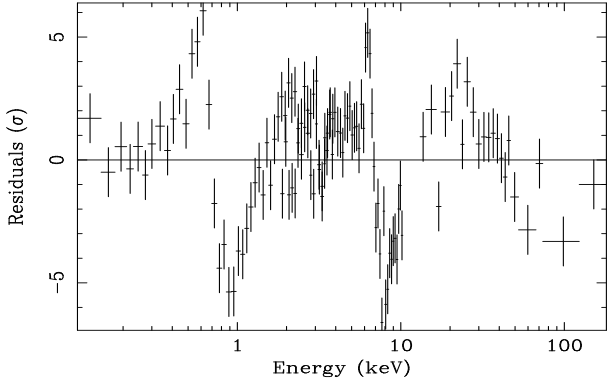}
\end{center}
\caption{Residuals in units of standard
deviation when a photoelectric absorbed power-law is
applied to the LECS, MECS and PDS data}
\label{fig1}
\end{figure}

We have fitted the spectra of all detectors simultaneously.
In Fig.~\ref{fig1} the residuals are shown, when a simple absorbed power law
is applied. The spectral
complexity is apparent: an edge-like feature at $\simeq 0.7$~keV is
the imprinting of warm absorbing matter along the line of sight
(Nandra \& Pounds 1992; Fabian et al. 1994);
an emission line, peaking at $\simeq 6.3$~keV,
is broadly consistent with K$_{\alpha}$ fluorescence from neutral or
mildly ionized iron; a continuum ``bump'', peaking at $\simeq 20$~keV,
is associated with Compton scattering of the primary continuum by
optically thick matter surrounding the central black hole. The last two
ingredients are known to be ubiquitously present in the X-ray spectra of
Seyfert 1 galaxies (Nandra \&
Pounds 1994; Matt 1998).

We have assumed therefore a ``baseline'' model composed by:
a power-law, a Compton-reflection component (model {\verb!pexrav!} in
{\sc Xspec}, Magdziarz \& Zdziarski 1995) with the inclination angle
held fixed to 35$^{\circ}$ (see later) and a ``warm absorber''. Following
Orr et al. (1997), the
warm absorber has been parameterized as a set of
four absorption edges, with threshold energies fixed at those expected
from the following ionized species: O{\sc vii}, O{\sc viii}, Ne{\sc ix}, Ne{\sc x}.
However, we were forced to add several further component to this
parameterization, in order to achieve a robust estimate of
the intrinsic power-law and a $\chi^2$ reasonably close to one, and
namely: a) a narrow emission line with centroid energy
${\rm E_C = 0.62 \pm^{0.04}_{0.03}}$~keV and Equivalent Width ${\rm EW = 33 \pm^{20}_{15}}$~eV;
b) a narrow emission line with
${\rm E_C = 0.86 \pm^{0.08}_{0.03}}$~keV and 
${\rm EW = 60 \pm^{20}_{40}}$~eV; c) an absorption edge with
threshold energy ${\rm E_{th} = 7.6 \pm^{0.3}_{0.2}}$~keV
and optical depth ${\rm \tau = 0.14 \pm^{0.03}_{0.05}}$.
These features are likely to be associated with the ionized absorber;
they are consistent with K$_{\alpha}$ O{\sc vii}, iron-L
(or  K$_{\alpha}$ Ne{\sc vii}--{\sc ix}) fluorescence transitions, and
photoabsorption from Fe{\sc xv}$\pm${\sc iv}, respectively.
A detailed discussion of these features is beyond the focus of this
{\it Letter}. Similar results on the
high-energy continuum and iron line as those
presented later are obtained if the self-consistent
warm absorber model {\verb!absori!} in {\sc Xspec} is used.

\begin{figure*}[hbt]
\begin{center}
\epsfig{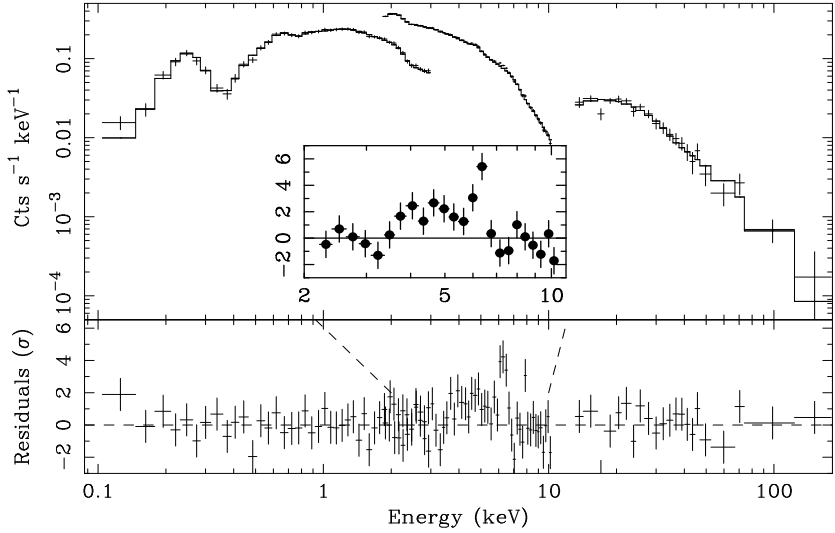}
\end{center}
\caption{Spectra ({\it upper panel}) and residuals in units of standard
deviation ({\it lower panel}) when the best-fit baseline continuum is
applied on the whole band except the 4--7.5~keV energy range (details in text).
In the {\it inset}, the MECS residuals only, binned by a further factor of
2. The best-fit parameters for such a continuum are: ${\Gamma = 2.00}$, ${\rm R = 1.2}$ and ${\rm E_{cutoff} = 130}$~keV}
\label{fig4}
\end{figure*}

In Fig.~\ref{fig4}, the residuals
are shown when the ``baseline'' model is applied to the whole
energy range, but excluding the 4--7.5~keV interval, where the
relativistic iron
line is expected to be present.
The iron line profile is {\it remarkably similar to that of ASCA}
(cf. Fig.~1 in Tanaka et al. 1995).
This is the first independent
confirmation of the existence of extreme relativistic effects
in the X-ray spectrum of MCG-6-30-15. This result is of particular importance
because of the more reliable continuum determination provided by BeppoSAX.

\begin{figure}
\begin{center}
\epsfig{figure=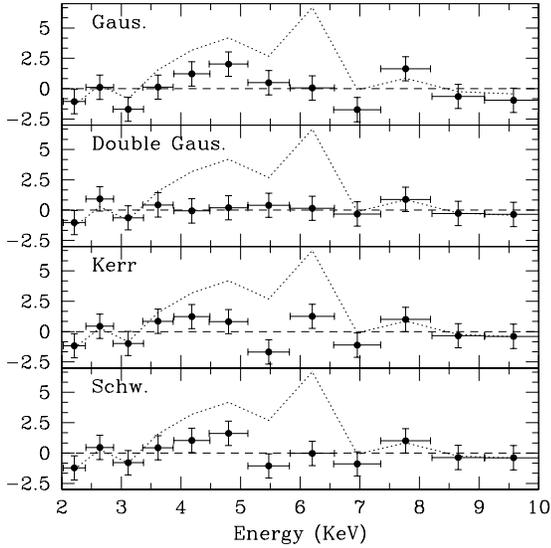,height=8.0cm,width=8.0cm}
\end{center}
\caption{MECS residuals in units of standard deviations when the
following models are employed to fit the iron line profile
(from top to bottom): single Gaussian, double Gaussian, Kerr relativistic
profile, Schwarzschild relativistic profile. In all panels, the
dotted line represents the envelope of the residuals in the inset
of Fig.~2, to illustrate, for sake of comparison,
the unfitted line profile against the underling continuum}
\label{fig2}
\end{figure}

Several descriptions of the line profile have been tried, whose
best-fit parameters are listed in Table~\ref{tab1}.
\begin{table*}
\caption{Best-fit results and parameters when the ``baseline'' continuum
is applied and the iron line modeled as in Column~1.}
\begin{center}
\begin{tabular}{lccccccccc} \hline
Model & ${\rm N_H}$ & ${\Gamma}$ & ${\rm R}$ & ${\rm E_{cutoff}}$ & ${\rm E_C}$ or ${\rm R_o}$ & ${\sigma}$ or ${\rm i}$& ${\rm EW}$  & ${\rm \chi^2}$ \\
& ($10^{20}$~cm$^{-2}$) & & & (keV) & (keV) or (${\rm r_g}$) & (keV) or $^{\circ}$ & (keV) & \\ \hline
Single Gaussian &  $6.3 \pm^{0.3}_{0.4}$ & $1.97\pm^{0.05}_{0.04}$ & $1.0 \pm 0.3$ & $110 \pm^{70}_{30}$ & $6.37 \pm 0.09$ & $< 0.18$ &  $50 \pm^{16}_{15}$ &  144.3/124 \\
Double Gaussian$^{\ddag}$ &  $6.7 \pm 0.4$ & $2.04 \pm 0.06$ & $1.3 \pm 0.4$ & $130 \pm 40$ & $4.5 \pm^{0.4}_{0.8}$ & $0.8\pm^{0.7}_{0.3}$ & $90 \pm^{230}_{40}$ and  &  130.6/123 \\
& & & & & & & $70\pm^{30}_{20}$$^{\flat}$ & \\
Schwarzschild & $6.8 \pm 0.3$ & $2.06 \pm 0.03$ & $1.2 \pm^{0.4}_{0.2}$ & $160 \pm^{130}_{60}$ & $7.2 \pm^{2.6}_{1.1}$ & $36 \pm 2$ & $200 \pm^{50}_{60}$ & 138.0/124 \\
Kerr & $6.8 \pm^{0.3}_{0.2}$ & $2.068 \pm^{0.025}_{0.018}$ & $1.2 \pm^{0.4}_{0.2}$ & $150 \pm^{110}_{50}$ & $11 \pm^6_5$  & $34 \pm^8_4$ & $220 \pm^{80}_{70}$ & 145.7/124 \\ \hline \hline
\end{tabular}
\end{center}
\noindent
$^{\dag}$fixed \\
$^{\ddag}$in this model, the second Gaussian line has ${\rm E_C = 6.4}$~keV and $\sigma = 0$. \\
$^{\flat}$for the broad and narrow components, respectively
\label{tab1}
\end{table*}
A simple Gaussian profile leaves significant residuals in the 4--5~keV
energy band (see Fig.~\ref{fig2}), which justify the addition of a second
broad line. If we assume that the first line is narrow and ``neutral''
(i.e.: centroid held fixed at 6.4~keV), ${\rm \Delta \chi^2}$ is equal to 13.7
for 1 further degree of freedom, significant at more than 99.9\% level.
However,
there is no straightforward physical explanations for a broad line with
${\rm E_C \simeq 4.5}$~keV
outside the relativistic scenario. 
We have then tried models of relativistic line
profiles around a Schwarzschild ({\verb!diskline!} in {\sc Xspec},
Fabian et al. 1989)
or a Kerr ({\verb!laor!} in {\sc Xspec}, Laor 1991) black hole. In both these
models the internal radius of the emitting region has been held fixed to the
last stable orbit (6 and 1.23 gravitational radii ${\rm r_g}$, respectively)
and the emissivity radial
function parameterized as ${\rm r^{-2}}$ (the results are marginally
affected by any choice of the emissivity index in the
range -2 to -3). The addition of a
further narrow Gaussian line to the relativistic models
yields a negligible change in the $\chi^2$
and the 90\%
upper limit on its EW is 40~eV. All the above models yield acceptable
fits at 90\% level.
However, even the best relativistic model is still not capable to
fully account for the redmost tail of the profile (see Fig.~\ref{fig2}).
Iwasawa et al. (1996) have shown that the contrast
between the blue horn and the red tail diminishes in very low intensity states.
MCG-6-30-15 exhibited variability by a factor of four on timescales
as low as a few thousands seconds during the BeppoSAX observation
(Orr et al. 1997). It is then likely that
the true line profile is a superposition of different states and
is then still different from the steady state template.
Common features of both the relativistic models
are an EW of the order of 200~eV, a rather small outer radius
of the line emitting region (${\rm R_o \rm \simlt 17}$ gravitational radii,
${\rm r_g}$) and an inclination
angle $\simeq$35$^{\circ}$.
A worse fit is obtained
in the Schwarzschild scenario (${\rm \chi^2 = 145.2/124}$~dof) if one
assumes a large outer radius ({\it e.g.}: ${\rm R_o = 10^3 r_g}$)
and allows the inner radius to be free.

\begin{figure*}
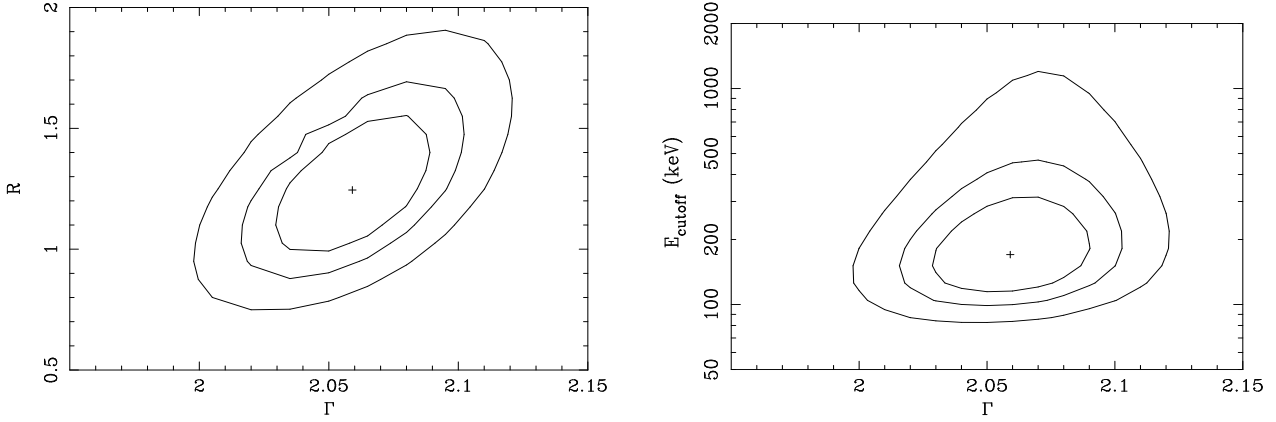

\begin{center}
\epsfig{figure=fig4b.ps,height=8.0cm,width=5.5cm,angle=-90}
\hspace{0.5cm}
\epsfig{figure=fig4a.ps,height=8.0cm,width=5.5cm,angle=-90}
\end{center}
\caption{{\it Left panel}: ${\rm \Gamma}$ vs.
${\rm R}$ contour plot for the baseline +
Schwarzschild relativistic line profile model.
{\it Right panel}: cutoff
energy vs. ${\rm \Gamma}$}
\label{fig3}
\end{figure*}

The best-fit broadband continuum parameters are in good agreement
with the ones observed in MCG-6-30-15
by previous missions (Fiore et al. 1992).
MCG-6-30-15 confirms to have a slightly steeper intrinsic continuum
($\Gamma \simeq 2.06$)
than the average measured in Seyfert~1 as a class
(Nandra et al. 1994; Nandra et al. 1997).
The amount of reflection (parameterized thorough
the relative normalization between the reflected and
the primary components ${\rm R}$) is consistent
with that expected from an
isotropically-illuminated, plane-parallel infinite slab. For the
first time, a cutoff in the primary continuum has been measured
in MCG-6-30-15, with
${\rm E_{cutoff}}$ comprised in the range 100--400~keV at 90\% confidence
level for two interesting parameters (see Fig.~\ref{fig3}).
The average observed 2--10~keV flux is $5.4 \times 10^{-11}$~erg~cm$^{-2}$~s$^{-1}$,
corresponding to an unabsorbed rest frame luminosity of $1.50 \times
10^{43}$~erg~s$^{-1}$ (we assume ${\rm H_0 = 50}$~km~s$^{-1}$~Mpc$^{-1}$
and ${\rm z = 0.008}$).
It is about 60\% higher than in the long-look ASCA observation
(Tanaka et al. 1995).

\section{Conclusions}

The analysis of the BeppoSAX observation of the Seyfert~1
galaxy MCG-6-30-15 fully confirms the ASCA discovery of a relativistic
iron line. The unprecedented broad bandwidth of BeppoSAX permitted a precise
determination of
the underlying continuum, thereby overcoming a potential criticisms to the
ASCA result. The amount of reflection continuum
is consistent
with the one assumed by Tanaka et al. (1995).
The bulk of the iron line is produced
very close to the black hole and the system is seen at a moderate inclination
angle, in agreement with ASCA findings.
However, the EW ($\simeq 200$~eV) measured by BeppoSAX is consistent with the
expected value for solar abundances, given the value of ${\rm R}$
(Matt et al. 1992). These data do not require any {\it ad hoc} iron
overabundance, such as the ASCA measurement of the same quantity
(${\rm \simeq 330}$~eV) had suggested (Tanaka et al. 1995).

\begin{acknowledgements}

The BeppoSAX satellite is a joint Italian--Dutch program.
MG and AO acknowledge the receipt of an ESA Research Fellowship.
GM acknowledges financial support from ASI and MURST.

\end{acknowledgements}

\end{document}